\documentclass{article}
\usepackage{cite}
\usepackage{amsmath,amssymb,amsfonts}
\usepackage{algorithmic}
\usepackage{graphicx}
\usepackage[nolist,nohyperlinks]{acronym}
\usepackage[utf8]{inputenc}
\usepackage{url}
\usepackage{color}
\usepackage{soul}
\usepackage[table,xcdraw]{xcolor}
\usepackage[caption=false, labelformat=simple]{subfig}
\usepackage{orcidlink}
\usepackage{tabularx}
\usepackage{pifont}
\usepackage{authblk}
\usepackage{etoolbox}

\newenvironment{keywords}{%
    \vspace{0.5em}%
    \noindent\textbf{Keywords:} %
}{%
    \par\vspace{0.5em}%
}

\newcommand\blfootnote[1]{%
  \begingroup
  \renewcommand\thefootnote{}\footnote{#1}%
  \addtocounter{footnote}{-1}%
  \endgroup
}

\newcommand{\xmark}{\ding{55}}

\hypersetup{
    pdfborder={0 0 0}
}

\newcommand{\twinningMethod}{m}
\newcommand{\setOfTwinningMethods}{\cal{M}}
\newcommand{\twinningParameter}{\theta}
\newcommand{\setOfTwinningParameters}{\Theta}
\newcommand{\episode}{e}
\newcommand{\totaleps}{E}
\newcommand{\step}{s}
\newcommand{\totalstepsperepisode}{S}
\newcommand{\totalsteps}{T}
\newcommand{\action}{a}
\newcommand{\RSS}{\textrm{RSS}}
\newcommand{\RSSestimation}{\widehat{\textrm{RSS}}}
\newcommand{\cost}{C}
\newcommand{\performance}{P}
\newcommand{\energyconsumption}{W}
\newcommand{\cumulativeenergyconsumption}{J}
\newcommand{\performanceweight}{w_{p}}
\newcommand{\energyweight}{w_{j}}

\newcommand{\precoder}{\textbf{f}}  
\newcommand{\channel}{\textbf{H}} 

\DeclareMathOperator*{\argmax}{arg\,max}
\newcommand{\codewords}{I}
\newcommand{\receivingsignal}{y}
\newcommand{\beampairindex}{i}

\begin{document}

\begin{acronym}
    \acro{MCA}{Mixed-Strength Covering Array}
    \acro{CPS}{Cyber-Physical System}
    \acro{IoV}{Internet of Vehicle}
    \acro{DL}{deep learning}
    \acro{VUE}{vehicle user}
    \acro{O-RAN}{Open RAN}
    \acro{3GPP}{Third Generation Partnership Project}
    \acro{ETSI}{European Telecommunications Standards Institute}
    \acro{ZSM}{Zero-touch Network and Service Management}
    \acro{ITU-T}{International Telecommunication Union Telecommunication Standardization Sector}
    \acro{IETF}{Internet Engineering Task Force}
    \acro{E2E}{end-to-end}
    \acro{mmWave}{millimeter wave}
    \acro{6G}{sixth generation of cellular networks}
    \acro{TPE}{Tree-structured Parzen Estimator}
    \acro{BO}{Bayesian Optimization}
    \acro{UPA}{Uniform Planar Array}
    \acro{RSS}{Received Signal Strength}
    \acro{RSSI}{Received Signal Strength Indicator}
    \acro{NDT}{Network Digital Twin}
    \acro{DTN}{{Digital Twin Network}}
    \acro{DT}{{Digital Twin}}
    \acro{PTwin}{{Physical Twin}}
    \acro{VTwin}{{Virtual Twin}}
    \acro{ML}{{Machine Learning}}
    \acro{MIMO}{{Multiple Input Multiple Output}}
    \acro{MISO}{{Multiple Input Single Output}}
    \acro{AI}{{Artificial Intelligence}}
    \acro{KPI}{{Key Performance Indicator}}
    \acro{RT}{{ray tracing}}
    \acro{Rx}{{receiver}}
    \acro{Tx}{{transmitter}}
\end{acronym}

\title{Energy-Driven Evaluation of Network Digital Twinning Applied to mmWave
  Beam Management}

\author[1]{João~Borges~\orcidlink{0000-0003-0038-8693}}
\author[1]{Bruno~Castro~\orcidlink{0000-0003-4601-3205}}
\author[2]{Kleber~Cardoso~\orcidlink{0000-0001-5152-5323}}
\author[3]{Andrey~Silva~\orcidlink{0000-0001-8320-8234}}
\author[1]{Aldebaro~Klautau~\orcidlink{0000-0001-7773-2080}} \affil[1]{LASSE -
  5G and IoT Research Group, Federal University of Pará (UFPA), Belém 66075-750,
  Brazil}
\affil[2]{Universidade Federal de Goiás, Goiânia 74001-970, Brazil}
\affil[3]{Ericsson Research, Rodovia Engenheiro Ermênio de Oliveira Penteado, Indaiatuba 13337-300, Brazil}

\date{}

\maketitle

\blfootnote{This work has been submitted to the IEEE for possible publication. Copyright may be transferred without notice, after which this version may no longer be accessible.}

\begin{abstract}
  Network Digital Twins (NDTs) are important enablers of 6G and future networks.
  However, there is a lack of studies regarding practical aspects, such as the
  impact of simultaneously changing twinning rate, fidelity, and other NDT
  operational parameters. For instance, works often consider the impact of
  operational parameters in isolation or with physical twin (PTwin)
  implementations relying on simulations. Therefore, the main contribution of
  this work is to provide an in-depth study on the performance impacts of both
  twinning rate and fidelity, using PTwins that rely on measurements obtained
  from hardware. We also investigate the optimization of these two operational
  parameters to minimize energy consumption, exploring a Bayesian Optimization
  (BO) method and what-if analysis. In this work, the NDT models an indoor
  propagation environment to optimize beam management. The virtual twin (VTwin)
  was implemented with the Sionna ray tracing (RT) simulator and the PTwin with
  our in-house setup composed of customized Wi-Fi radios with 32 antenna
  elements operating at 60 GHz. The study reveals important aspects of wireless
  channel NDTs, suggesting that fidelity levels vary throughout the experiment
  and with the what-if difficulty. Moreover, the twinning rate can be adjusted
  using sample-efficient methods, such as BO, even at lower fidelity levels.
\end{abstract}


\section{Introduction}
\label{sec:intro}
The so-called \ac{DT} systems are currently envisioned as key
enablers of the \ac{6G}~\cite{tran2025network, khan2022digital}. Among its main
features are the capacity of performing actions such as real-time system
management, as well as a process known as \textit{what-if analysis}, where a
hypothetical scenario is first tested in a virtual replica before deployment in
the real system \cite{3gpp_tr32801_01, ibm2024dtwhatif}. These features prospect
the existence of a safe test ground for network rollouts, expansions, and
automations, so to support them, the \ac{DT} systems exploit heavy integration
with sophisticated simulations and \ac{AI} \cite{sheraz2024comprehensive}.
However, as the reliance on these technologies grows, so do the concerns with
resource consumption \cite{sheraz2024comprehensive} \cite{liu2025dtmeets}. For
use cases such as \ac{MIMO} beam management, this is further enforced due to a
current lack of works focused on measuring and optimizing the resources employed
to enable \ac{NDT} operation. This work explores the performance impact of
varying the \emph{twinning rate} and \emph{fidelity}, which are two of the most
characteristic \ac{DT} parameters and part of the \ac{DT} definition according
to the Digital Twin Consortium \cite{dtconsortium2026definition}, with the goal
of saving energy resources without compromising the system performance. Some
related works focus on developing state-of-the-art methods to increase
fidelity and maintain synchronization, e.g. \cite{khalaf2026uavaidedDTSynch},
which is out of the scope of this work. The purpose of this paper is to evaluate
the effects of simultaneously varying \ac{NDT} parameters and checking their
accuracy vs energy consumption performance. The problem is posed as a
sequential discrete-continuous decision problem with episodic feedback, where we
check the effects on the system performance when using different twinning rate
policies present in the literature and their respective configurations, under
two different fidelity levels, first using an ideal replica and then an
imperfect one. For the twinning rate policies choice and configuration we use
two approaches, a grid-search method, using a limited set of configurations, and
a \ac{BO} method, for a more sample-efficient parameter space exploration.
Besides these, we also consider a ``max-rate" and a ``no-retwinning'' policies
for baseline purposes, the first which performs twinning on every step and the
second that only performs twinning in the first step and do not execute it
again. For the performance evaluation, we use a composite metric which considers
both twinning error and energy consumption. Tests are carried out on an in-house
\ac{NDT} system, where the \ac{PTwin} is composed of customized Wi-Fi radios
with 32 antenna elements operating at 60 GHz and the \ac{VTwin} by the NVIDIA
Sionna \ac{RT} simulator \cite{hoydis2023sionnart}.

The main contributions of this paper are:
\begin{itemize}
  \item the investigation of the simultaneous effect of two main \ac{NDT}
        operational parameters on the system performance;
  \item the exploration of a Bayesian method for sample-efficient \ac{NDT}
        twinning method and parameter selection;
  \item the introduction of a standards-based \ac{NDT} architecture for the beam selection, described in
        Section~\ref{sec:ndt_system}; and
  \item a hardware implemented \ac{PTwin}, described in Section~\ref{sec:evaluation}.
\end{itemize}

The remaining content of this paper is structured as follows:
Section~\ref{sec:related_work} navigates the \ac{DT}/\ac{NDT} terminology and
core concepts necessary for the paper, as well as the related work in \ac{DT}
operation, with emphasis on twinning. Then,
Section~\ref{sec:problem_formulation} addresses the mathematical definition of
the energy-driven evaluation problem for the \ac{NDT} twinning and \ac{MISO}
beam selection. Next, on Section~\ref{sec:ndt_system} we explain the \ac{NDT}
architecture as well as the what-if process pipeline we follow. In
Section~\ref{sec:evaluation}, the twinning rate and fidelity experiments, as
well as the
\ac{NDT} setup considered in this work are described, and then
discussed. Finally, Section~\ref{sec:conclusion} concludes the paper,
summarizing the contributions and proposing future works.

\section{Background and Related Work}
\label{sec:related_work}

\renewcommand{\arraystretch}{1.5}
\begin{table}[htb]
  \centering
  \caption{Related work}
  \smallskip
  \resizebox{\linewidth}{!}{
    \begin{tabular}{|c|c|c|c|c|c|}
      \hline
      \rowcolor[HTML]{C0C0C0}
      \textbf{Work}                                                           &
      \textbf{\begin{tabular}[c]{@{}c@{}}Twin rate\\ evaluation\end{tabular}} &
      \textbf{\begin{tabular}[c]{@{}c@{}}Twin fidelity\\ evaluation\end{tabular}}
                                                                              & \textbf{\begin{tabular}[c]{@{}c@{}}PTwin uses physical \\ measurements
                                                                                          \end{tabular}} & \textbf{NDT}                 & \textbf{Use
      case}                                                                                                                                                                                                                                                                                          \\ \hline
      Muñoz et al., 2022 \cite{munoz2022usingtrace}                           & \textcolor{red}{\xmark}                                                & \textcolor{blue}{\checkmark} & \textcolor{blue}{\checkmark} & \textcolor{red}{\xmark}      & DT fidelity evaluation in CPSs                 \\ \hline
      Tan et al., 2022 \cite{tan2022optimizingDTsynch}                        & \textcolor{blue}{\checkmark}                                           & \textcolor{red}{\xmark}      & \textcolor{red}{\xmark}      & \textcolor{red}{\xmark}      & Optimizing DT synchronization in manufacturing \\ \hline
      Cakir et al., 2023 \cite{cakir2023synchdt}                              & \textcolor{blue}{\checkmark}                                           & \textcolor{red}{\xmark}      & \textcolor{red}{\xmark}      & \textcolor{blue}{\checkmark} & Network infrastructure for DT synchronization  \\ \hline
      Zheng et al., 2023 \cite{zheng2023datasync}                             & \textcolor{blue}{\checkmark}                                           & \textcolor{red}{\xmark}      & \textcolor{red}{\xmark}      & \textcolor{blue}{\checkmark} & Data synchronization in vehicular networks     \\ \hline
      Tan et al., 2024 \cite{tan2024theDTsynchProblem}                        & \textcolor{blue}{\checkmark}                                           & \textcolor{red}{\xmark}      & \textcolor{red}{\xmark}      & \textcolor{red}{\xmark}      & Optimizing DT synchronization in manufacturing \\ \hline
      Muñoz et al., 2024 \cite{munoz2024measuring}                            & \textcolor{red}{\xmark}                                                & \textcolor{blue}{\checkmark} & \textcolor{blue}{\checkmark} & \textcolor{red}{\xmark}      & DT fidelity evaluation in CPSs                 \\ \hline
      Salehi et al., 2024 \cite{salehi2024multiverse}                         & \textcolor{red}{\xmark}                                                & \textcolor{blue}{\checkmark} & \textcolor{blue}{\checkmark} & \textcolor{blue}{\checkmark} & MIMO beam selection in V2X scenarios           \\ \hline
      Huang et al., 2025 \cite{huang2025beam}                                 & \textcolor{red}{\xmark}                                                & \textcolor{blue}{\checkmark} & \textcolor{red}{\xmark}      & \textcolor{blue}{\checkmark} & MIMO beam selection in V2X scenarios           \\ \hline
      Our work                                                                & \textcolor{blue}{\checkmark}                                           & \textcolor{blue}{\checkmark} & \textcolor{blue}{\checkmark} & \textcolor{blue}{\checkmark} & MISO beam selection in indoor scenarios        \\ \hline
    \end{tabular}
  }
  \label{tab:related_works}
\end{table}

In this section, we provide a brief overview of core concepts useful for the
literature analysis, such as the difference between \acp{DT} and \acp{NDT}, as
well as an explanation about the twinning rate and fidelity, which are the
\ac{DT} parameters explored in this investigation. Finally, at the end of the
section, we provide insights into the current related works studying the individual
effects of these concepts over \acp{DT} and \acp{NDT} systems and list this work
main differences.

\subsection{Digital Twins and Network Digital Twins}

The term \ac{DT} has a widely accepted origin as part of the work of Michael
Grieves and John Vickers \cite{grieves2014digital, jones2020characterisingDT}.
According to their early description, the digital
twin is a virtual representation of a physical target, containing real
information and consisting of three elements: the physical target
itself, its virtual representation, and the bi-directional data flow
between them. \acp{DT} are able to perform tasks such as system state monitoring, logging and, using \textit{what-if} scenarios it can preemptively test changes on the system e.g. upgrades and downgrades, as well as run feature tests or check its behavior in alternative scenarios, all these in the virtual replica, without risking the real system \cite{ibm2024dtwhatif}.

Besides \ac{DT}, the more specific term \ac{NDT} is used when it comes to the
communication networks. Here we adopt the \ac{3GPP} definition of \acp{NDT}
\cite{3gpp_tr28915}, which is also shared by the \ac{ETSI} \ac{ZSM}
\cite{ETSI-2024-NDT}, and the \ac{IETF} \cite{IETF-2025-arch_proposal}, where an
\ac{NDT} is a \ac{DT} in which the replicated element is the network itself, or just part of it. From this definition, we can infer that \acp{DT} of
\ac{E2E} networks, single domains, or even just the communications channel, or
the user device, are all \acp{NDT}.

There is also some disambiguation to be made regarding the naming convention.
For instance, the \ac{O-RAN} \cite{ORAN-2024-DTRAN_usecases} and also the
\ac{ITU-T} \cite{ITU-2022-DTN_reqs_and_arch} both refer to \acp{NDT} by the term
\ac{DTN}.
Alternatively, some works use \ac{DTN} to define a network of many
single \acp{DT}~\cite{Wu-2021-DT_Networks}.

\subsection{Types of Twinning Rates}
\label{subsec:types_twinning_rates}
The twinning rate is defined in \cite{jones2020characterisingDT} as the
frequency that the virtual replica updates its states based on the twinned
target. In other words, it is the rate in which the synchronization between \ac{VTwin}
and \ac{PTwin} occurs. The authors in \cite{alghamdi2024synchronization} provide
$11$ synchronization patterns for general \acp{DT}, which are named
event-driven, time-driven, hybrid, adaptive, multi-level, event-chain,
data-driven, context-aware, multi-modal, asynchronous, and manual. From these,
we consider three patterns, regarded as within the scope of the
setup
and experiments used in this work, and provide practical implementations of
them. They are: the event-driven, time-driven, and adaptive. The first pattern,
event-driven is further divided into two variants, the threshold-based and the
event-based ones, meant to cover metric monitoring events and also more
intermittent triggers, respectively. For their practical implementation, the
first variant, named threshold-based, defines a given numeric \ac{KPI} to
monitor, which in this paper is the system \ac{RSS}, and it monitors whether it
falls beyond a given dBm value. The second variant, identified as event-based,
defines a list of $N$ events, not linked to the numeric \ac{KPI} monitoring,
where the twinning will occur. From now on, we refer to these two variants
(\textit{threshold-based} and \textit{event-based}) as their own synchronization patterns for
easier identification of the twinning method used in the experiments. Besides
these, the next pattern, i.e., the time-driven, employs as trigger of the twinning
process a fixed interval of time. Finally, the last one, i.e., the
adaptive synchronization pattern, uses both an
interval of time and a threshold, similar to the time-driven and
threshold-based modes. The difference is that the time interval is variable, and also, twinnings triggered by trespassing the threshold take precedence over
those triggered by time. More specifically, in this work, a given initial
interval and threshold are defined and followed for a given amount of time. Once
this time is finished, the interval is incremented, and the process continues as
long as the monitored metric stays within the threshold. However, if the threshold is crossed, the twinning is executed, and the interval is sent back to the initial
value.

\subsection{Twin Fidelity Performance Metric}
The term \textit{fidelity} appears throughout \ac{DT} literature
\cite{zhang2018equipment, luo2019digital, zhuang2018digital,
  zheng2019application, guo2019modular}, with \cite{jones2020characterisingDT}
defining the fidelity of a \ac{VTwin} as the abstraction level, the set of
parameters used in the replica, and its actual accuracy in representing the
\ac{PTwin}. In this work, we follow this definition, however, when considering
different fidelities in our experiments, we only vary the actual accuracy of
representing the target, and do not focus on the abstraction level and set of
parameters used in the representation.

Some works evaluate digital twin fidelities by comparing the sequence of
system states \cite{munoz2022usingtrace, munoz2024measuring}. We perform
a similar evaluation, but on a higher level, by comparing the results from the top-K
chosen beams. This is due to the lack of lower-level information
availability in the physical measurements of the dataset.

Regarding the accuracy metric used for this evaluation, we adopt the \\
``$precision@k$", which is defined in~\cite{charchar2025}:
\begin{equation}
  \text{precision@}k =
  \frac{|{\cal{I}}_v \cap {\cal{I}}_p|}{k},
  \label{eq:precisionk}
\end{equation}
where ${\cal{I}}_v$ is the set of best indices suggested by the \ac{VTwin} and
${\cal{I}}_p$ is the set of best indices from \ac{PTwin} measurements. As
mentioned in~\cite{charchar2025}, the term \textit{precision} references the common
\ac{ML} performance measure with the same name, which is the ratio of
\textit{true positives} to the total number of \textit{predicted positives},
both true and false ones. For the $precision@k$, we consider the \textit{true
  positives} as $|{\cal{I}}_v \cap {\cal{I}}_p|$, which represents the \ac{VTwin}
correctly guessing the beam indices that compose the Top-K, and the total amount
of \textit{predicted positives} as $k$, given that the beam indices outside the
${\cal{I}}_p$ are \textit{predicted negatives}.

\begin{figure}[t]
  \centerline{\includegraphics[trim=0.7cm 0.45cm 0.7cm 0.4cm, clip, width=\textwidth]{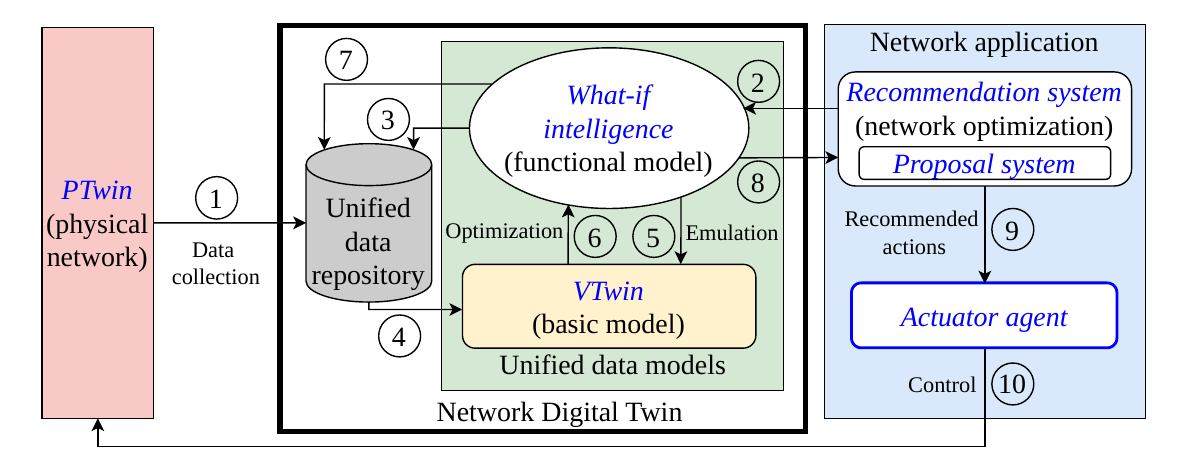}}
  \caption{Overview of the system architecture, with elements and interactions inspired by the ones from the ITU-T. The elements and/or names in both italic and blue represent those that are not present and/or are named differently in the ITU-T architecture.}
  \label{fig:system_overview}
\end{figure}

\subsection{Related Work}
The investigation of \ac{DT} systems is extensive due to its pervasiveness,
being present on several industries, such as robotics, aerospace, manufacturing,
transportation, agriculture, and healthcare, to name a few
\cite{alzami2025DTindustriesSurvey}. However, despite the large number of works,
the majority of these are focused on investigating the benefits the \ac{DT}
brings to the system where it is installed, but do not consider the impacts of
different deployment parameters on the use case performance, nor the \ac{DT}'s cost of operation. Those that do so,
deal with them separately, for instance, with a focus either on the twinning rate or
the virtual replica fidelity, but not on two or more parameters simultaneously.
The number of works further reduces when it comes to the specific case of
\acp{NDT} and when the implementation of the \ac{PTwin} in hardware is
considered, as some works target only the \ac{VTwin}, without any physical measurements.
Also, in the current literature, there is a frequent assumption of the ideal case
for the \ac{VTwin}. But a virtual replica can rarely achieve 100\% perfect
mimicking, and this non-ideal condition affects the search for an optimal
twinning rate, increasing the relevance of considering both the twinning rate and
fidelity during the investigations.

In Table~\ref{tab:related_works}, we list the works that most closely align
with ours when it comes to the presence of the \ac{DT} parameters studied, as
well as the presence of an actual hardware implementation of the \ac{PTwin}, and
also with the indication of it being an \ac{NDT} and their target use case. In
\cite{munoz2022usingtrace} and \cite{munoz2024measuring}, Muñoz et al.
investigate methods of evaluating the fidelity level of \ac{DT} of \acp{CPS}.
These works focus on comparing traces of robotic systems' operations to assess
the real and virtual counterparts' alignment quality. In
\cite{tan2022optimizingDTsynch} and \cite{tan2024theDTsynchProblem}, Tan et al.
explore the search for an optimal twin synchronization for general \acp{DT}
using mathematical formulations, but with a fixed \ac{DT} fidelity and without a
hardware implementation. In \cite{cakir2023synchdt}, Cakir et al. investigate
how the communication system configuration, alongside the desired twinning rate,
affect both, the synchronization performance between \ac{PTwin} and \ac{VTwin},
determining the maximum achievable twinning rate, and also the use case
performance metric, but without addressing variations in twinning fidelity. In
\cite{zheng2023datasync}, Zheng et al. focus on the synchronization of
\acp{VUE} and the \ac{DT} in an \ac{IoV} setting by using a \ac{DL}-based network
delay prediction approach, while also not considering variations on fidelity or
the usage of measurements from real hardware as the \ac{PTwin}. In
\cite{salehi2024multiverse}, Salehi et al. leverage the FLASH dataset
\cite{salehi2022FLASH} as the \ac{PTwin} and explore the usage of multiple
\acp{VTwin} with different fidelity levels to balance beam selection performance
gains and the what-if analysis latencies associated with each of them, but do
not explore the twinning rate. In \cite{huang2025beam}, Huang et al. leverage
knowledge about the scenario to optimize the tradeoff between system complexity
and its performance. This is done by varying the number of antennas and,
therefore, the \ac{VTwin} fidelity, and performing a scenario
cognition-empowered beam selection that balances performance and system
complexity, without relying solely on maximum fidelity. They also, however, do
not explore the variation of a second parameter, namely, the twinning rate.

Finally, to the best of the authors' knowledge, there is currently no prior work
investigating the energy-versus-performance impact of varying multiple \ac{DT}
parameters (i.e., twinning rate and fidelity) while considering a \ac{DT} for
communications network, in other words, an \ac{NDT}, involving both, a hardware
implemented \ac{PTwin} and a \ac{VTwin}, applied for the \ac{mmWave} beam
selection problem.

\section{Problem formulation}
\label{sec:problem_formulation}
In this section, we provide the mathematical definition for both the energy-driven twinning evaluation and the \ac{MISO} beam selection problem. We start with basic terminology definitions and proceed to explain the system evolution with the formulas of the monitored performance and energy metrics, then we explain the optimization function and why we can use a \ac{BO} approach as a solution.

\subsection{Basic definitions}
\label{subsec:basic_definitions}

The \ac{NDT} system management considered in this work is described as a dynamic
decision making problem where the \ac{NDT} system must choose between a twinning
method $\twinningMethod \in \setOfTwinningMethods$ from a set of $\setOfTwinningMethods$ possible twinning
methods, and then its respective parameters $\twinningParameter_\twinningMethod \in
  \setOfTwinningParameters_\twinningMethod$, from a set $\setOfTwinningParameters_\twinningMethod$ of parameters, to reduce both, the twinning
error and the system energy consumption. This choice occurs at the start of an
\emph{episode} $\episode$ from a total of $\totaleps$ episodes, where $\episode =
  1,..., \totaleps$, each occurring over a series of steps $\step$, with $\step = 1, ...,
  \totalstepsperepisode$, considering $\totalstepsperepisode$ describing the total
amount of steps needed for the completion of an episode, so that for all
episodes we have a total number of steps $\totalsteps$, where $\totalsteps = \totaleps \times
  \totalstepsperepisode$. This series of choices is summarized in the tuple
$\action$, which is an action that must be taken at the start of each episode
$\episode$, so that we have $\action_\episode = (\twinningMethod_\episode, \twinningParameter_{\twinningMethod_{\episode}})$.

\subsection{System evolution}
\label{subsec:system_evolution}
The system evolves in episodes, each ending with a given cost $\cost$, that
considers both, the cumulative twinning performance $\performance$ and the
cumulative energy consumption $\cumulativeenergyconsumption$. The performance
$\performance$ is the cumulative absolute error between the measured \ac{RSS}
and its estimation ($\RSSestimation$), while $\cumulativeenergyconsumption$ is
the sum of the energy consumption $\energyconsumption$ in Watt-hour of
performing the twinning process followed by what-if analysis, both over the
entire episode $\episode$, and subject to the chosen action $a_\episode$.
Finally, the performance $\performance$ and the cumulative energy consumption
$\cumulativeenergyconsumption$ of the episode $\episode$ using the action
$\action_\episode$ are respectively defined as
\begin{equation}
  \performance_\episode(\action_\episode) = \sum_{\step=1}^{\totalstepsperepisode} |\RSS_{\step} - \RSSestimation_{\step}| \ ; \ \ \ \forall \ \episode \in \totaleps,
  \label{eq:cumulative_twin_error}
\end{equation}
and
\begin{equation}
  \cumulativeenergyconsumption_\episode(\action_\episode) = \sum_{\step=1}^{\totalstepsperepisode} \energyconsumption_{\step} \ ; \ \ \ \forall \ \episode \in \totaleps.
  \label{eq:cumulative_energy_usage}
\end{equation}

Due to their different value ranges, these two metrics have their samples
normalized using min-max scaling, bringing their values to the $[0,1]$ interval,
and with their normalized versions receiving the notation $\tilde{\performance}$
and $\tilde{\cumulativeenergyconsumption}$. Finally, we can say that the cost
$\cost$ at the end of the episode $\episode$, using the method
$\twinningMethod_\episode$ and parameters
$\twinningParameter_{\twinningMethod_{\episode}}$ is
\begin{equation}
  \cost_\episode(\twinningMethod_\episode, \twinningParameter_{\twinningMethod_{\episode}}) = \cost_\episode(\action_\episode) = \performanceweight \times \tilde{\performance}_{\episode}(\action_\episode) + \energyweight \times \tilde{\cumulativeenergyconsumption}_{\episode}(\action_\episode),
  \label{eq:cumulative_cost}
\end{equation}
where $\performanceweight$ and $\energyweight$ are complementary scalar weights
within the range $[0,1]$. These were added to adjust the focus on one of the two
metrics, depending on the system optimization priority, namely the twinning error and
the energy expenditure, respectively. For this work we use the values
$\performanceweight = \energyweight = 0.5$.

\subsection{Optimization function and solution approach}
\label{subsec:optimization}

Considering the described dynamic decision-making problem, we have the following
equation as the target
\begin{equation}
  \hat{\action}(\hat{\twinningMethod}, \hat{\twinningParameter}_\twinningMethod) = \underset{\twinningMethod, \twinningParameter_\twinningMethod}{\operatorname{argmax}} \mathop{\mathbb{E}}[\cost_\episode(\twinningMethod_\episode, \twinningParameter_{\twinningMethod_{\episode}})] \ \ \forall \ \episode \in \totaleps,
  \label{eq:optimal_method_and_parameter_search}
\end{equation}
which represents the search for the optimal action $\hat{\action}$, described in
terms of the optimal twinning method $\hat{\twinningMethod}$ and its respective
optimal parameters $\hat{\twinningParameter}_\twinningMethod$. These values enable
the system to achieve the optimal cost by minimizing the expected twinning error
and the energy expenditure of each episode.

Due to the nature of this problem, the method chosen for this search
leverages a \ac{BO} method, more specifically, the \ac{TPE}
\cite{bergstra2011algorithms}, because of its capabilities of accommodating
categorical variables such as the ones from the discrete twinning method choice,
and also dealing with hierarchical sequential decision problem, with
discrete-continuous decision and episodic feedback, such as the choice of
twinning rate method, and then the search for its parameters.

\begin{figure}[htb]    \centerline{\includegraphics[width=1.07\linewidth]{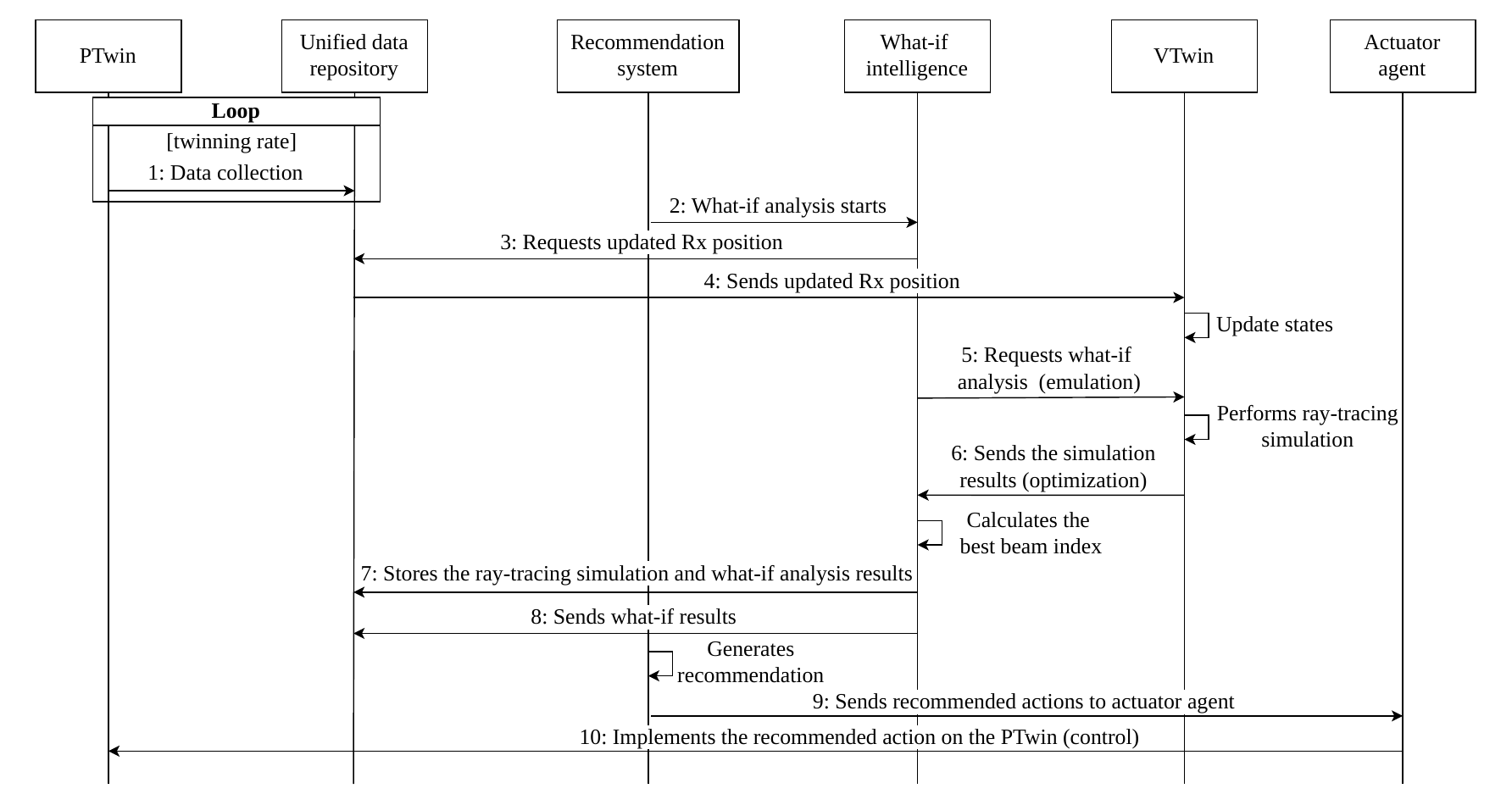}}
  \caption{Sequence diagram of a what-if scenario generation using the proposed ITU-T inspired NDT system architecture.}
  \label{fig:seq_diag}
\end{figure}

\subsection{The MISO beam selection process}
\label{subsec:beam_selection}
In this work, we consider the \ac{MISO} beam selection problem to be described by
a scenario with a \ac{Tx} and \ac{Rx} \acp{UPA}, with precoder represented by
the vector $\precoder$, which is used alongside the channel matrix $\channel$ to
obtain the receiving signal $\receivingsignal$ for a given beam pair index
$\beampairindex$. In this scenario, the receiving antenna uses only one element,
therefore, the receiving signal $\receivingsignal$, given the index
$\beampairindex$, is calculated by
\begin{equation}
  \receivingsignal_\beampairindex = \channel\precoder_\beampairindex,
  \label{eq:receiving_signal}
\end{equation}
and the optimal beam choice $\hat{\beampairindex}$ is given by
\begin{equation}
  \hat{\beampairindex} = \argmax_{\beampairindex \in {1,..., \codewords}} |\receivingsignal_\beampairindex|.
  \label{eq:bestbeam}
\end{equation}

\section{NDT System}
\label{sec:ndt_system}

In this section, we present the \ac{NDT} architecture considered for the
experiments and explore the what-if scenario generation. The architecture is
inspired by the one from Y.3090 \cite{itu2022y3090}, defined by the ITU-T and
shown in Fig.~\ref{fig:system_overview}, where the names in black correspond to
terminology and elements directly from the ITU-T architecture, while the ones in
blue represent alternative names or added elements. Also, we explain how they
interact with each other during a what-if scenario implementation via a
sequence diagram presented in Fig.~\ref{fig:seq_diag}.

This what-if scenario considers the beam selection problem described in
Subsection~\ref{subsec:beam_selection}. As illustrated in
Fig.~\ref{fig:system_overview} and Fig.~\ref{fig:seq_diag}, we have the
\textit{\ac{PTwin}}, also named physical network in the system architecture,
which is composed by a scenario with a fixed and a mobile \ac{UPA}, that can be
either the \ac{Tx} or the \ac{Rx}, depending on the simulation settings. The
position of this mobile \ac{UPA} is captured at a given frequency, defined by
the \textit{twinning rate} parameter, and logged at an element named
\textit{unified data repository} in the \ac{NDT} system, which in this case is a
database. This is the \textbf{step 1}, denoted as \textit{data collection}. Once
reaching the \textit{unified data repository}, the \ac{Rx} position data is
considered to be at disposal of the \ac{NDT}. This data collection step is a
loop and keeps repeating until a what-if analysis is initiated by the
\textit{recommendation system} in \textbf{step 2}. Once started, the
\textit{what-if intelligence} requests updated \ac{Rx} position information from
the \textit{unified data repository} during \textbf{step 3}, that are then
transmitted to the \ac{VTwin} at \textbf{step 4}, which updates its internal
states with the newest \ac{Rx} position.

\begin{figure*}[t]
  \centering
  \subfloat[Top view of the physical area covered in this work for the in-house NDT setup, obtained by a camera mounted on the roof.\label{fig:inhouse_ptwin_topview}]{%
    \includegraphics[trim=1cm 0cm 6cm 1cm, clip, width=.475\textwidth]{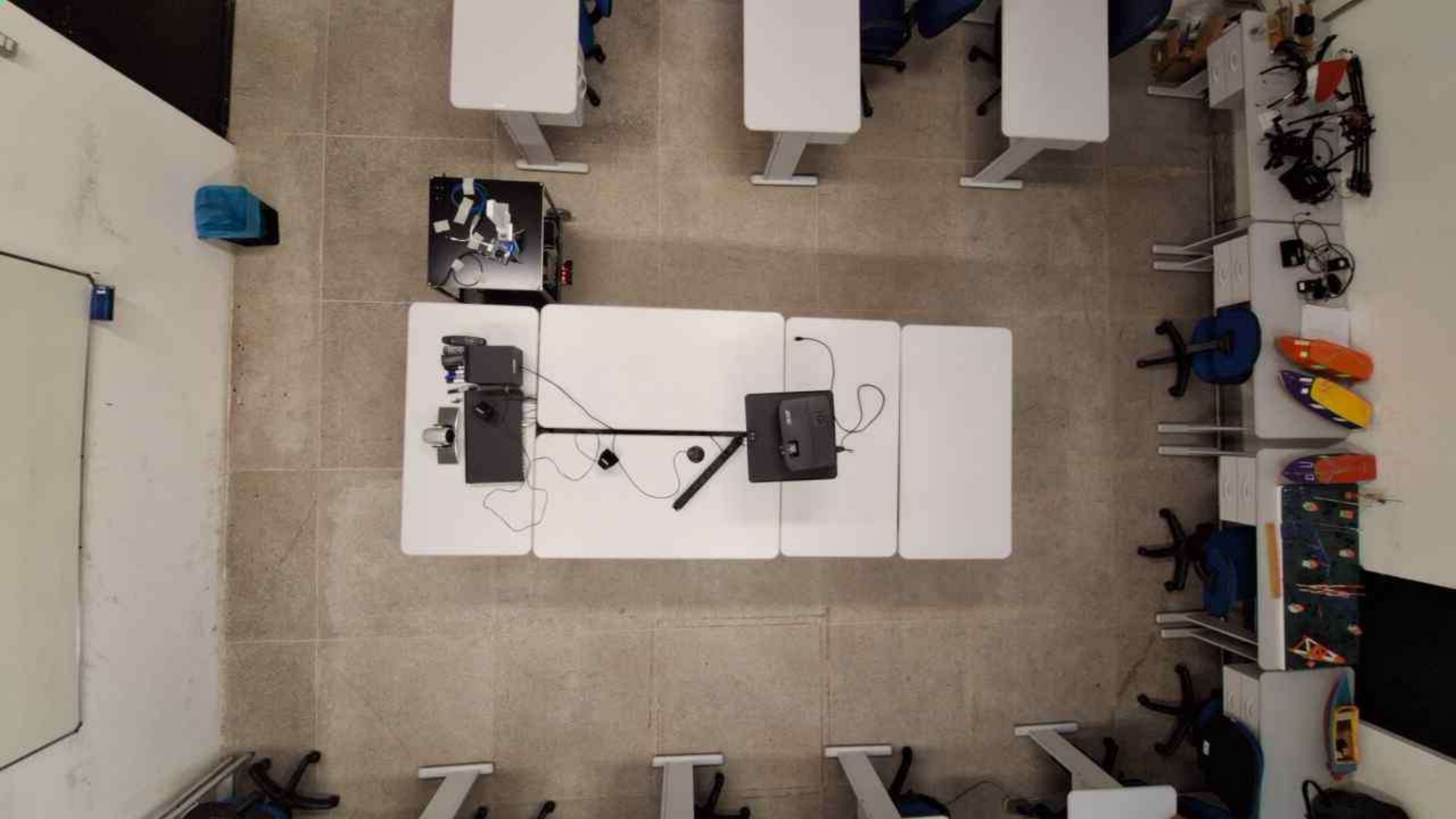}%
  }
  \hfill
  \subfloat[Top view of the room virtual model for the in-house NDT setup, generated using the open-source software Blender.\label{fig:inhouse_vtwin_topview}]{%
    \includegraphics[trim=1cm 1.5cm 2cm 4cm, clip, width=.4\textwidth]{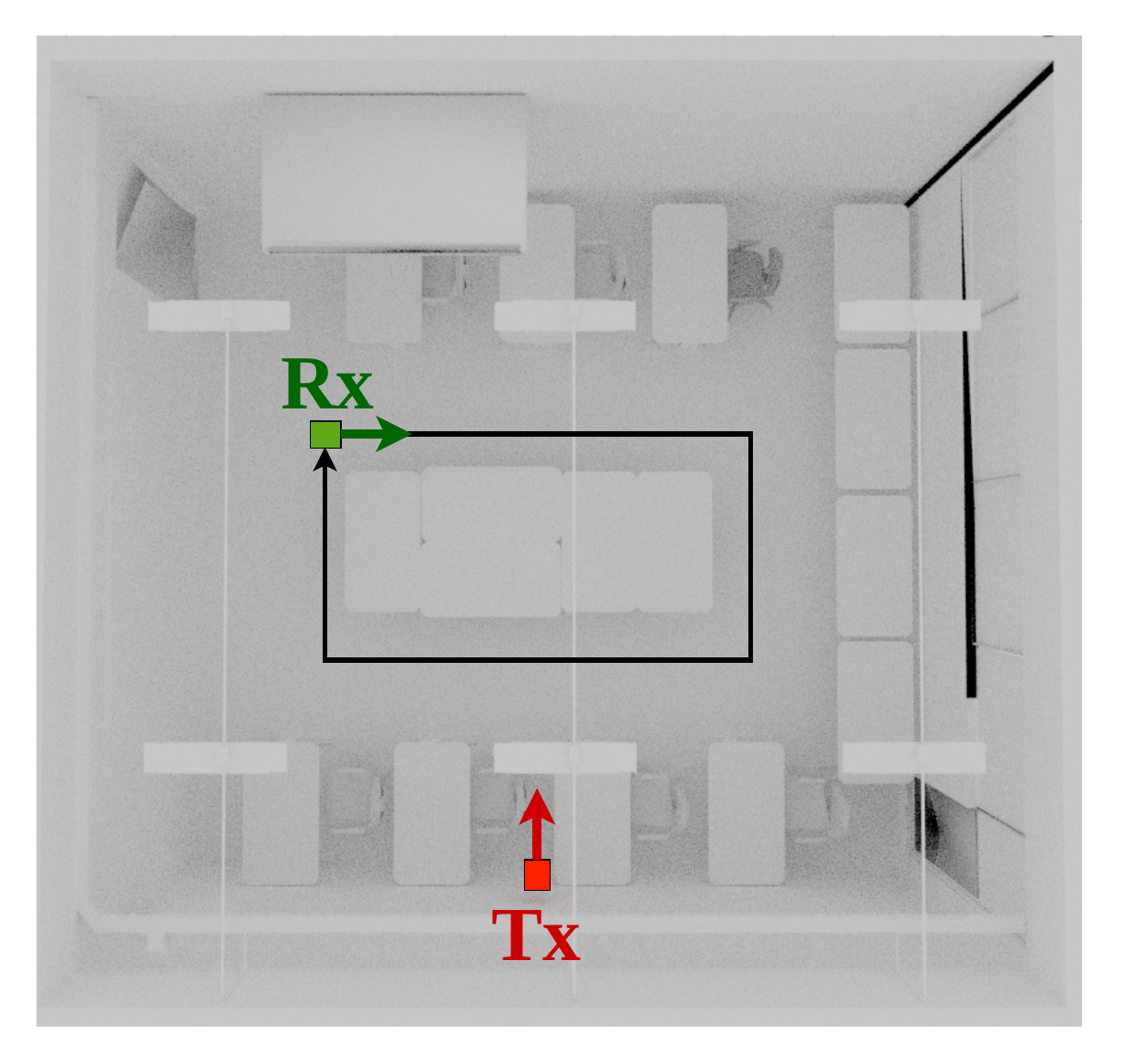}%
  }
  \caption{Real and virtual versions of the room used in the in-house NDT setup experiments. (a) Physical area: a laboratory room with the main elements being a whiteboard, side and center tables, and a robot carrying the MikroTik setup, which obtained the RSSIs; (b) virtual model: generated using the open-source software Blender. The figure also highlights the placement of the Tx (red), the Rx (green), and the path followed by the robot (black line). The red arrow indicates the orientation of the Tx, while the Rx always points to the direction of the robot movement around the center table.}
  \label{fig:inhouse_ptwin_and_vtwin_topviews}
\end{figure*}

The \ac{NDT} is composed by a \textit{unified data repository} and a
\textit{unified data model}, which in turn is composed of \textit{\ac{VTwin}}
and \textit{what-if intelligence}, also known as \textit{basic model} and
\textit{functional model}, respectively. The \textit{\ac{VTwin}} is a virtual
representation of the \textit{\ac{PTwin}}, described with a given fidelity,
while the \textit{what-if intelligence} is the element capable of extracting
useful insights from this representation by simulating/emulating proposed
actions, e.g., to optimize the system. This interplay between the
\textit{\ac{VTwin}} and the \textit{what-if intelligence} is described in
both, the Y.3090 and our architecture, in terms of \textit{emulation} and
\textit{optimization}, identified by the \textbf{step 5} and \textbf{step 6}.
In \textbf{step 5}, the \textit{what-if intelligence} requests an analysis to
check the best beam index (emulation), for which the \ac{VTwin} performs a
\ac{RT} procedure and in \textbf{step 6}, it sends the \ac{RT} data back to
the \textit{what-if intelligence}, that calculates the best beam index
according to the received data (optimization). Then, the \ac{RT} data,
alongside the what-if results, are propagated to the \textit{unified data
  repository} for logging purposes, and only the what-if results to the
\textit{recommendation system}, in \textbf{steps 7} and \textbf{8},
respectively.

In \textbf{step 9}, the \textit{recommendation system}, via the
\textit{proposal system}, generates the recommendation, consisting of the best
index for the \ac{Tx}, and sends to the \textit{actuator agent}. Finally, in
\textbf{step 10}, the \textit{actuator agent} implements it in the
\textit{\ac{PTwin}} (control).

\section{Evaluation}
\label{sec:evaluation}
In this section we explain the twin fidelity and twinning rate analysis experiments as well as the experimental setup used for them. At the end, we provide the results and discussion regarding the interplay between twinning rate and fidelity in regards to the adopted energy-aware performance metric.

\subsection{Analysis on the Twin Fidelity and the Twinning Rate}
\label{subsec:twinning_fidelity_and_rate_experiments}
For the experiments of the twin fidelity impact on the system performance, we
execute the beam selection procedure on the \ac{VTwin} and compare with the
ground-truth results. The experiment consists of evaluating the $precision@k$
throughout the path covered by the mobile \ac{UPA}, as shown in
Fig.~\ref{fig:inhouse_vtwin_topview}. The experiment is composed of $47$
episodes, each with $58$ steps, where for every step we have a top-K for both,
\ac{PTwin} and \ac{VTwin}, so that we can calculate the $precision@k$ of all the
steps for every episode.

Concerning the twinning rate, we are evaluating the
variation on the cumulative cost, integrating twinning error and energy
expenditure as defined in Eq.~\ref{eq:cumulative_cost}. In the experiments, we
consider a receiver moving in the indoor scenario shown in
Fig.~\ref{fig:inhouse_ptwin_topview}. The \ac{NDT} system has at its disposal a
model-driven \ac{VTwin} based on ray tracing simulation using NVIDIA Sionna
\cite{sionna} for channel modeling. The tests are carried out on a single test
episode, while the other episodes are used for parameter search using the
\ac{TPE}.

\begin{table}[htb]
  \caption{Simulation Parameters Adopted for the VTwins}
  \centering
  \begin{tabular}{lc}
    \hline
    \multicolumn{2}{c}{\textbf{Simulation parameters}}                       \\ \hline
    \textbf{Sionna version}                & 1.1.0                           \\
    \textbf{Material database}             & ITU                             \\
    \textbf{Transmitter}                   & UPA 6x6                         \\
    \textbf{Receiver}                      & Single element                  \\
    \textbf{Frequency/bandwidth}           & 60 GHz                          \\
    \textbf{Antenna pattern (Tx/Rx)}       & TR.38901                        \\
    \textbf{Antenna polarization}          & Vertical                        \\
    \textbf{Max. num. of ray interactions} & 6                               \\
    \textbf{Max. num. of paths per Tx}     & $10^6$ (to allow virtually all) \\
    \textbf{Num. of rays launched per Tx}  & $10^6$                          \\
    \textbf{Synthetic array}               & True                            \\
    \textbf{Allow LOS paths}               & Enabled                         \\
    \textbf{Specular reflection}           & Enabled                         \\
    \textbf{Diffuse reflection}            & Enabled                         \\
    \textbf{Refraction}                    & Enabled
  \end{tabular}
  \label{tab:vtwin_parameters}
\end{table}

We follow the process described in Fig.~\ref{fig:seq_diag}, where the system
initially stays in a loop, feeding the \textit{unified data repository} with
up-to-date receiver positions, as described in \textbf{step 1}. However, differently from
the default behavior shown there, once the experiment starts, each twinning is
immediately followed by the trigger of a what-if analysis. This what-if consists
of obtaining the beam index with higher \ac{RSS} from a codebook of
$64$ possible \ac{Tx} beams. The \ac{Rx} adopts a single receiving beam, with a
directional pattern informed in Table~\ref{tab:vtwin_parameters}. The \ac{Rx}
always points to the direction of the robot movement around the center table.
The process then follows through with \textbf{steps 2} to \textbf{10}.

For this experiment, we first perform an initial twinning at the first step and
then, for the retwinning policy, we contrast the use of the twinning rates
mentioned in Section~\ref{subsec:types_twinning_rates}, i.e., 1) time-driven, 2)
threshold-based, 3) adaptive, and 4) event-based. More specifically, the
time-driven method consists of defining a fixed interval for the twinning to
occur, which for this experiment must assume the value of an integer within the
range $[2, 58]$. The threshold-based monitors if the \ac{RSS}
achieved at the latest step falls below a given float number, and if it does, a
twinning is triggered. The adaptive uses both an interval and a threshold, with
the difference that the interval is variable instead of fixed. It works by
following the initial interval two times, and if the \ac{RSS} stays within the
threshold during this period, then the interval is incremented by one.
Otherwise, if the \ac{RSS} falls below the threshold, then a twinning is
immediately triggered and the interval goes back to the initial value. This
behavior reduces the energy consumption of executing a twinning followed by a
what-if analysis if the performance is not degraded enough to call for a new
twinning. Finally, the event-based twinning rate works by defining a set
composed of $N$ events, which are integer numbers within the range $[1, 58]$,
representing the steps for twinning to occur.

\begin{figure*}[htb]
  \centering
  \subfloat[Fidelity boxplots assuming $K=16$.\label{fig:twin_fidelity_result_k16}]{%
    \includegraphics[width=.49\textwidth]{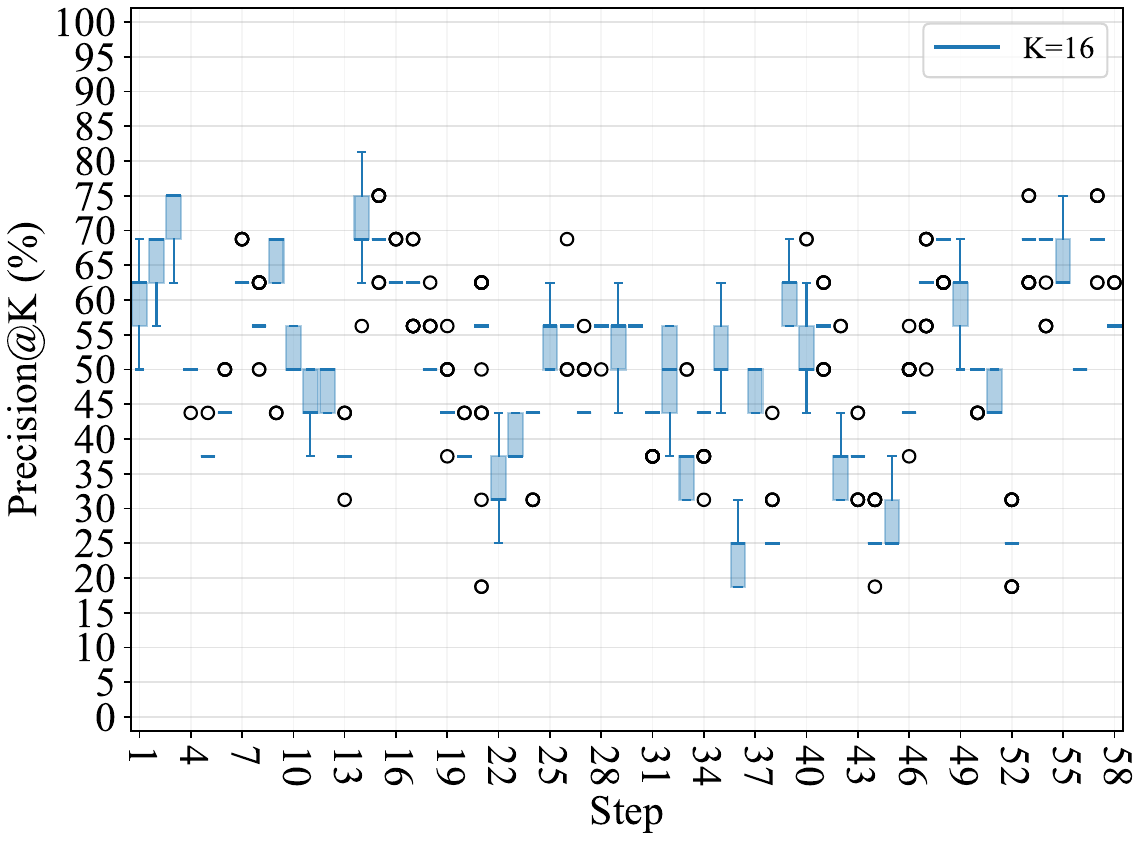}%
  }
  \hfill
  \subfloat[Fidelity boxplots assuming $K=32$.\label{fig:twin_fidelity_result_k32}]{%
    \includegraphics[width=.49\textwidth]{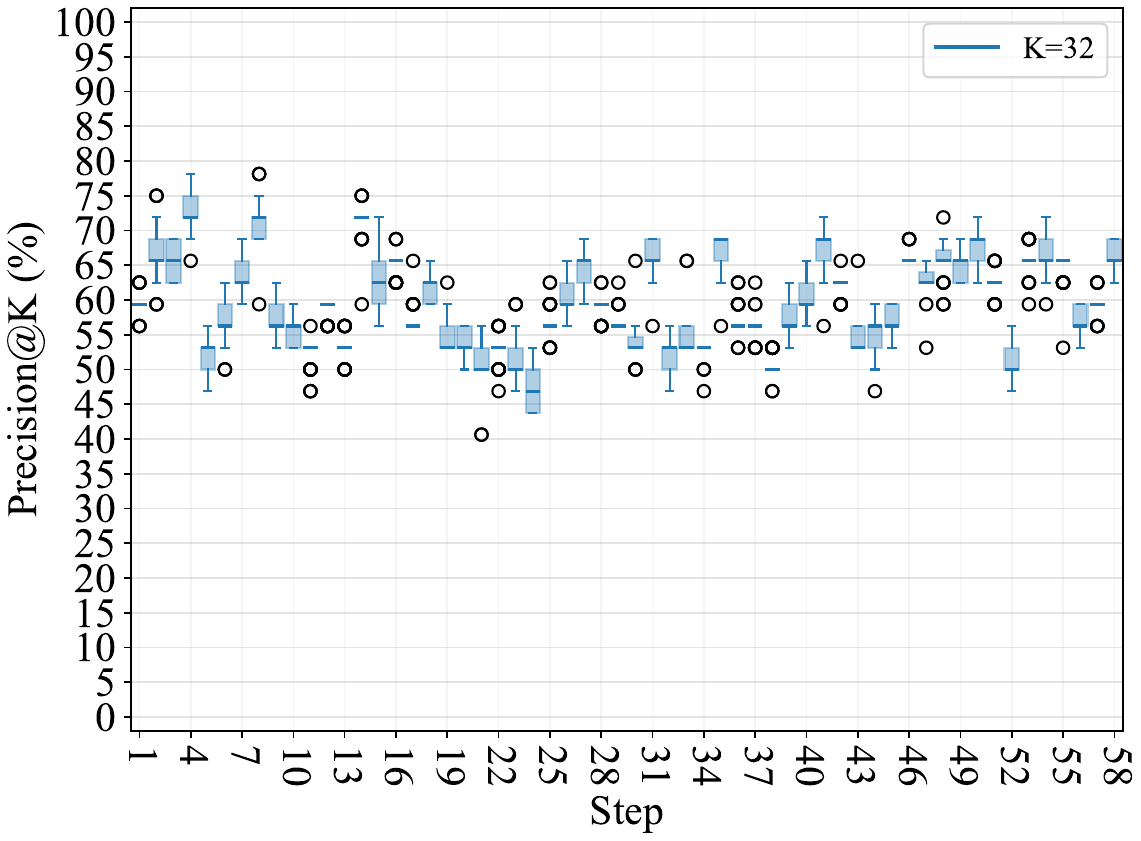}%
  }
  \caption{Boxplots of the fidelity at each step considering the entire dataset. The dots represent outliers, while the lower/upper whiskers are, respectively, the smallest/largest values besides the outliers. Using the ``$precision@k$" metric, the fidelity varies throughout the experiment evolution and with the chosen K. Lower K values suggest a more stringent what-if analysis, trying to pinpoint the best index, while higher K values widen the search to a bigger subset.}
  \label{fig:twin_fidelity_result}
\end{figure*}

The first four methods are initially implemented with parameters chosen using a
rudimentary grid-based hyperparameter search. After that, for a sample-efficient
search within the constrained dataset, a \ac{TPE} approach is used to find both,
a performative method and its respective hyperparameter(s). The grid search is
an exhaustive parameter sweep over a finite set of parameters that compose the
grid \cite{geron2022hands}. The parameters for it are predefined and chosen
empirically as described in the following:

\begin{itemize}
  \item \textbf{time-driven} -- the intervals can be a value from the set $\{2,
          4, 8, 10\}$, representing the steps. For instance, if $2$ is chosen, the
        twinning is executed every $2$ steps;
  \item \textbf{threshold-based} -- the threshold can assume one of the
        following values in dBm $\{-65, -60, -55, -50\}$;
  \item \textbf{adaptive} -- needs to define two parameters:
        \begin{itemize}
          \item \raggedright a threshold (in dBm) from the set $\{-65, -60, -55, -50\}$ and
          \item \raggedright the initial interval (in seconds) from the set $\{1, 2, 4, 8\}$;
        \end{itemize}
\end{itemize}

For the event-driven, aiming for better coverage of the search space we adopt
varying set sizes ($N=1, 2, 3$), with $5$ candidates for each. Also, the choice
of values to be swept by the grid-search is based on a Stratified Mixed-Strength
Design, a methodology inspired by \acp{MCA} \cite{2004ColbournCombinatorial,
  2012AvilaSimulated, 2006MooreCombinedArray} from the area of combinatorial
design and experimental optimization, used to uniformly partition the events
within the $[1, 58]$ value range. Therefore, the sets are the following:

\begin{itemize}
  \item \textbf{Single-Event Sets ($N=1$):} \\ $\{6\}, \{18\}, \{30\}, \{42\},
          \{54\}$.
  \item \textbf{Two-Event Sets ($N=2$):} \\ $\{3, 35\}, \{15, 47\}, \{22, 58\},
          \{10, 28\}, \{40, 51\}$.
  \item \textbf{Three-Event Sets ($N=3$):} \\ $\{2, 29, 57\}, \{12, 33, 49\}, \{7,
          20, 45\}, \{19, 38, 53\}, \\ \{5, 25, 41\}$.
\end{itemize}

Now, the \ac{TPE} is used to search for both, a twinning rate type and its
respective parameters, however, differently from the grid-search, the \ac{TPE}
can handle a much larger search space. It begins with a categorical search space
composed of \textit{time}, \textit{threshold}, \textit{adaptive} or
\textit{event}. After that, the search follows as described below:

\begin{itemize}
  \item \textbf{time-driven} -- the intervals can be an integer value from the
        range $[1:58]$ (steps);
  \item \textbf{threshold-based} -- the threshold can assume any float value
        within the range $[-80:0]$ (dBm);
  \item \textbf{adaptive} -- needs to define two parameters:
        \begin{itemize}
          \item a float threshold from the range $[-80:0]$ (in dBm) and
          \item an integer initial interval from the range $[1:58]$ (in steps);
        \end{itemize}
  \item \textbf{event-based} -- parameters choice is again, a set of size $N=1,
          2, 3$, where each element can assume integer values from the interval $[1:58]$,
        which represents the experiment ``episode" step.
\end{itemize}

In summary, for this experiment, the goal is to maximize the use case performance metric while at the same time reducing the number of times we trigger a twinning process request to the \ac{NDT}. In our case, this is immediately followed by a what-if scenario calculation, therefore also saving energy resources as reflected in the performance metric defined in Eq.~\ref{eq:cumulative_cost}.

\subsection{In-house experimental setup}
\label{subsec:inhouse_setup}
This work considers the indoor environment illustrated in
Fig.~\ref{fig:inhouse_ptwin_topview}. The data acquisition relies on a low-cost
experimental platform designed for \ac{AI}-based beam management research. The
setup consists of two MikroTik wAP 60G radios serving as the \ac{Tx} and
\ac{Rx}. They operate in the $60$ GHz mmWave band with $6\times6$ \ac{UPA}
antennas, having the four corners disabled and so, considering only $32$ active
elements. The \ac{Tx} is positioned next to a wall, by the side of the room,
while the \ac{Rx} is mounted on the top of a mobile robot, that follows a
circular path around the center table, with their placement shown in
Fig.~\ref{fig:inhouse_vtwin_topview}. The MikroTik radios are commercial
off-the-shelf devices, customized with an OpenWRT operating system and a
modified firmware for the IEEE 802.11ad baseband controller, enabling the
implementation of custom beamforming codebooks and the real-time extraction of
the \ac{RSSI} \cite{conceicao2025lowcost}.

In previous works, we developed a synchronized client-server system that manages
the entire data collection process, capturing time-aligned multimodal data that
combines wireless metrics (e.g., the \ac{RSSI}) with images from RGB and depth
cameras \cite{conceicao2025lowcost, valdinei2025repo}. To enable mobility, the
\ac{Rx} radio is mounted on a robotic platform programmed to autonomously follow
a predefined path in the environment. This integrated system allowed for the
creation of robust datasets that correlate wireless channel quality with visual
information in dynamic scenarios. Further details on the platform architecture
and its validation can be found in \cite{conceicao2025lowcost,
  valdinei2025repo}.

Also, in this work, we leveraged the digital model of the room developed
in~\cite{charchar2025}, where the experimental setup is located, shown in
Fig.~\ref{fig:inhouse_vtwin_topview}. But, while in \cite{charchar2025} we used
the proprietary software Remcom Wireless InSite \cite{WirelessInSiteWebSite},
here we use the open-source NVIDIA Sionna as the \ac{VTwin}.

Next, we focus on the description of the experiments for monitoring the impact of
varying these two parameters on the system performance.

\subsection{Results}
\label{subsec:results}
Results obtained from the in-house experimental setup for the \ac{VTwin} fidelity
assessment are illustrated in Fig.~\ref{fig:twin_fidelity_result}, where in
Fig~\ref{fig:twin_fidelity_result_k16} and
Fig~\ref{fig:twin_fidelity_result_k32} we show the fidelity levels achieved for
$precision@k=16$ and $precision@k=32$, respectively. One can notice that the fidelity
levels vary throughout the experiment evolution and is also heavily influenced
by the chosen $k$. This result shows that a \textit{high fidelity} can be achieved
partially, throughout the experiments and also when the task at hand involves
less accuracy. For instance, by not using $K=1$ and instead increasing it, we
are changing the what-if task from suggesting the single best beam index to just
narrowing down the suggested optimal beam index to a $K$-sized subset of the original
codebook.

\begin{figure*}[htb]
  \centering
  \subfloat[Using the ideal VTwin. \label{fig:twin_rate_ideal}]{%
    \includegraphics[trim={0.3cm 0.3cm 0.45cm 0.4cm}, clip, width=.49\textwidth]{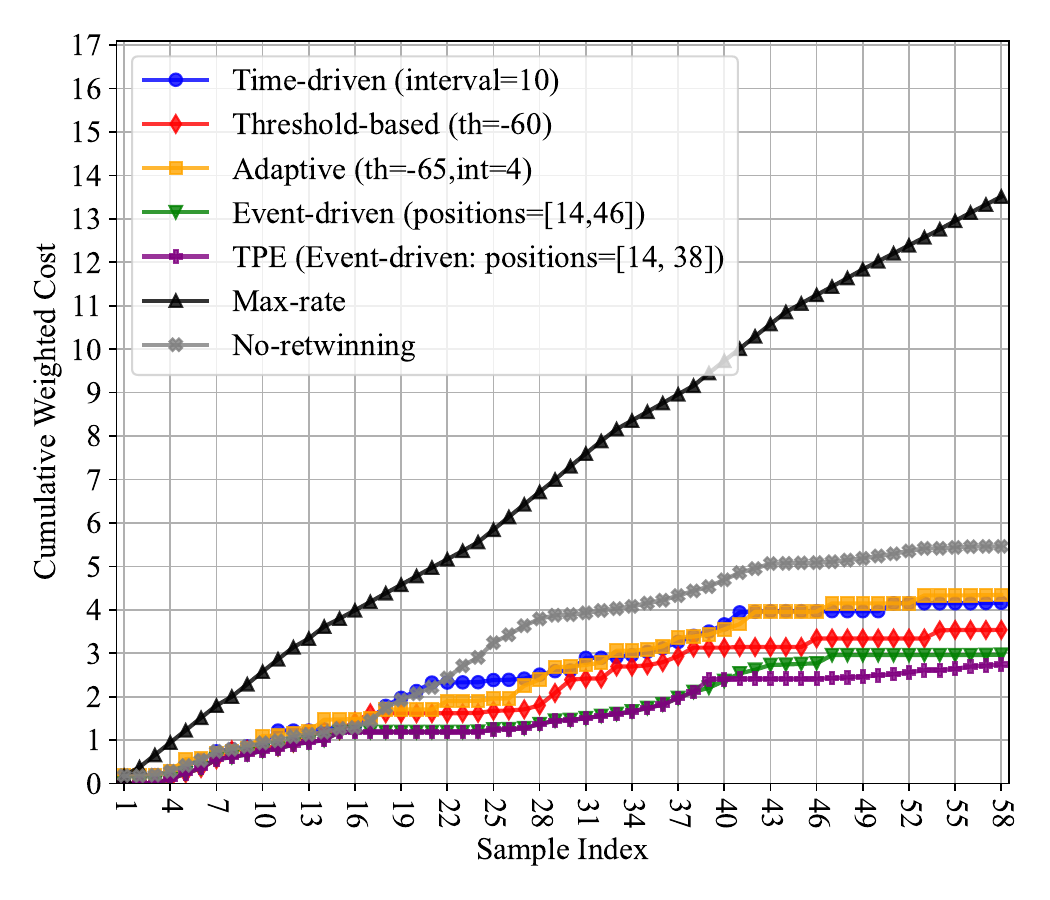}%
  }
  \hfill
  \subfloat[Using the Sionna VTwin. \label{fig:twin_rate_sionna}]{%
    \includegraphics[trim={0.3cm 0.3cm 0.45cm 0.4cm}, clip, width=.49\textwidth]{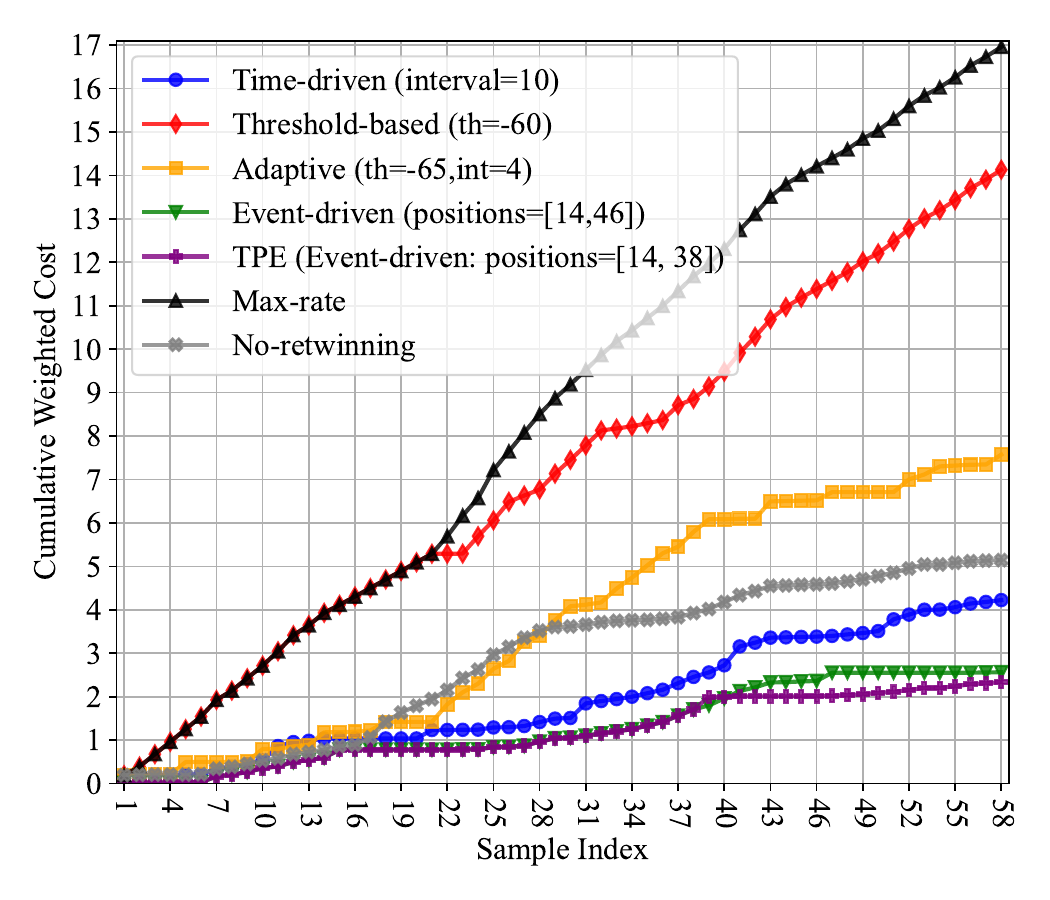}%
  }
  \caption{Cumulative weighted cost of both versions of the VTwin, the ``ideal'' one, that yields perfect fidelity, and the one using Sionna RT simulator, with varying fidelity.}
  \label{fig:twin_rate}
\end{figure*}

In Fig.~\ref{fig:twin_rate_ideal} and Fig.~\ref{fig:twin_rate_sionna}, we
present the variation of the cumulative cost defined in
Section~\ref{subsec:system_evolution} when using different types of twinning
rates and fidelities. In Fig.~\ref{fig:twin_rate_ideal} we explore the case
where the fidelity is ideal, i.e. the \ac{VTwin} is always at the highest
fidelity possible and we focus only on variations of the twinning rate, while in
Fig.~\ref{fig:twin_rate_sionna} we explore the case where the fidelity varies
throughout the experiment by using the Sionna-based \ac{VTwin}. In both figures
we compare the twinning methods described in \ref{subsec:types_twinning_rates},
as well as the max-rate, no-retwinning and the \ac{TPE} approaches. For the
twinning methods, we first execute a preliminary step, which is a grid search
for time-driven, threshold-based, adaptive, and event-driven. From this search,
we choose only the most performative parameters from each twinning method to be
compared with the remaining approaches. Next, the max-rate and the no-retwinning
are brought into comparison to provide the baseline for both, a system that is
constantly twinned, meaning that for every step it executes a new twinning, and
for a system that only performs twinning on the first step and then do not
perform it again. Lastly, the \ac{TPE} approach comes with a more sophisticated
search through the parameter space. Notice that, the less costly and therefore
best method in Fig.~\ref{fig:twin_rate_ideal}, \ac{TPE}, displays robustness
against performance degradations in Fig.~\ref{fig:twin_rate_sionna}, when the
\ac{VTwin} presents the non-ideal conditions shown in
Fig.~\ref{fig:twin_fidelity_result}.

\subsection{Discussion}
\label{subsec:discussion}
From the results we found that, even when
calibrated, \ac{VTwin} fidelity levels are usually dependent on the experiment time
evolution and the stringency of what-if conclusions. We also checked that a
correctly chosen twinning rate
method and respective parameters yield relevant gains in performance. To
provide insights into solving the latter, we proposed the usage of the \ac{TPE}
bayesian optimization method for a sample-efficient search within the available
candidates, which gave positive results.

We also observed that, the correct choice of twinning rate method and parameters
can greatly influence energy savings and high-performance maintenance. The
\ac{TPE} method, investigated as candidate solution for the twinning rate method
and respective parameters selection, provided the best results for the integrated
energy vs performance metric introduced in Section~\ref{sec:evaluation} when
compared with methods and parameters chosen using the grid-search method with
empirical value ranges, when used in both ideal and non-ideal fidelity
conditions, which was the case when using Sionna as the \ac{VTwin}.

Regarding the non-ideal fidelity observed on the Sionna-based \ac{VTwin}, two
possible causes are: simplified assumptions regarding the radiation patterns
adopted in the simulated environment antennas, and also, simplifications of the
room geometry itself, which for instance, do not capture the entire
complexity of the objects present in the room.

\section{Conclusion}
\label{sec:conclusion}
In this work, we investigated the tradeoff between \ac{VTwin} state update rate and
virtual to real accuracy, namely the twinning rate and fidelity, respectively.
By establishing an integrated comparison metric that allowed for energy-aware
performance comparison, we evaluated different twinning rates present in the
literature at different fidelity stages. We also investigated the usage of a
\ac{BO} approach, more specifically \ac{TPE}, to provide a sample-efficient
\ac{NDT} parameter optimization alternative. To achieve our results, we used
data involving both real measurements acting as the \ac{PTwin} and its
respective model-based \ac{VTwin}, leveraging a simulated environment and
\ac{RT} channel modelling. The results showed that the fidelity levels are
dependent on both the experiment time evolution and the objective of the
what-if. In other words, the fidelity will vary throughout the experiment, and
it will be more or less prominent if the goal of the what-if is to narrow down
the set of options for the suggested action, or to provide a single optimal
suggestion, respectively.

The findings represent a contribution to the investigations regarding the
\ac{NDT} deployment payoff and resource-aware operation, which tends to be of
critical importance as the technology adoption grows, given the projected
massive pervasiveness of \ac{6G} networks. Future work is to be focused on
expanding this investigation to more datasets and more sophisticated setups,
also exploring the viability of performing the same type of experiments for
other network domains or even \ac{E2E} networks.

\section*{Acknowledgements}
This study was financed in part by the Coordenação de Aperfeiçoamento de Pessoal
de Nível Superior - Brasil (CAPES) – Finance Code 001; in part by the
Conselho Nacional de Desenvolvimento Científico e Tecnológico (CNPq); in
part by the Brasil 6G project (01245.020548/2021-07), supported by Rede
Nacional de Ensino e Pesquisa (RNP) and Ministério da Ciência, Tecnologia e
Inovacão (MCTI); in part by the Innovation Center, Ericsson Telecomunicações
Ltda., Brazil; in part by the OpenRAN Brazil - Phase 2 project (MCTI grant
Nº A01245.014203/2021-14); and in part by the Project Smart 5G Core And MUltiRAn
Integration (SAMURAI) [MCTI/Comitê Gestor da Internet no Brasil
    (CGI.br)/Fundacão de Amparo à Pesquisa do Estado de São Paulo (FAPESP)]
under Grant 2020/05127-2

\bibliographystyle{IEEEtran}
\bibliography{references.bib}

@article{tran2025network,
  author   = {Tran, Dinh-Hieu and others},
  journal  = {IEEE Open Journal of the Communications Society},
  title    = {{Network Digital Twin for 6G and Beyond: An End-to-End View Across Multi-Domain Network Ecosystems}},
  year     = {2025},
  volume   = {6},
  number   = {},
  pages    = {6866-6911},
  doi      = {10.1109/OJCOMS.2025.3599866}
}

@article{khan2022digital,
  title     = {{Digital-twin-enabled 6G: Vision, architectural trends, and future directions}},
  author    = {Khan, Latif U and Saad, Walid and Niyato, Dusit and Han, Zhu and Hong, Choong Seon},
  journal   = {{IEEE Communications Magazine}},
  volume    = {60},
  number    = {1},
  pages     = {74--80},
  year      = {2022},
  publisher = {IEEE}
}

@techreport{3gpp_tr32801_01,
  author      = {{3rd Generation Partnership Project}},
  title       = {{Study on management and orchestration}},
  institution = {3GPP},
  type        = {Technical Report},
  number      = {TR 32.801-01},
  version     = {0.2.0},
  year        = {2026},
  month       = {April},
  url         = {https://www.3gpp.org/DynaReport/32801-01.htm}
}

@misc{ibm2024dtwhatif,
  author       = {{IBM Topics}},
  title        = {{What Is a Digital Twin?}},
  howpublished = {\url{https://www.ibm.com/think/topics/digital-twin}},
  year         = {2024},
  note         = {Accessed: 2026-05-17}
}

@article{sheraz2024comprehensive,
  author  = {Sheraz, Muhammad and Chuah, Teong Chee and Lee, Ying Loong and Alam, Muhammad Mahtab and Al-Habashna, Ala’a and Han, Zhu},
  journal = {IEEE Access},
  title   = {{A Comprehensive Survey on Revolutionizing Connectivity Through Artificial Intelligence-Enabled Digital Twin Network in 6G}},
  year    = {2024},
  volume  = {12},
  pages   = {49184-49215},
  doi     = {10.1109/ACCESS.2024.3384272}
}

@article{liu2025dtmeets,
  author  = {Liu, Wenshuai and Fu, Yaru and Shi, Zheng and Wang, Hong},
  journal = {IEEE Communications Magazine},
  title   = {{When Digital Twin Meets 6G: Concepts, Obstacles, and Research Prospects}},
  year    = {2025},
  volume  = {63},
  number  = {3},
  pages   = {16-22},
  doi     = {10.1109/MCOM.001.2400202}
}

@misc{dtconsortium2026definition,
  author       = {{Digital Twin Consortium}},
  title        = {{The Definition of a Digital Twin}},
  howpublished = {\url{https://www.digitaltwinconsortium.org/initiatives/the-definition-of-a-digital-twin/}},
  year         = {2026},
  note         = {Accessed: May 12, 2026}
}

@article{khalaf2026uavaidedDTSynch,
  author  = {Khalaf, Ghofran and Itani, May and Sharafeddine, Sanaa},
  journal = {IEEE Transactions on Network and Service Management},
  title   = {{A UAV-Aided Digital Twin Framework for IoT Networks With High Accuracy and Synchronization}},
  year    = {2026},
  volume  = {23},
  number  = {},
  pages   = {3013-3025},
  doi     = {10.1109/TNSM.2026.3670040}
}

@inproceedings{hoydis2023sionnart,
  author    = {Hoydis, Jakob and others},
  booktitle = {2023 IEEE Globecom Workshops (GC Wkshps)},
  title     = {{Sionna RT: Differentiable Ray Tracing for Radio Propagation Modeling}},
  year      = {2023},
  volume    = {},
  number    = {},
  pages     = {317-321},
  doi       = {10.1109/GCWkshps58843.2023.10465179}
}

@inproceedings{tan2022optimizingDTsynch,
  author    = {Tan, Bariş and Matta, Andrea},
  booktitle = {2022 Winter Simulation Conference (WSC)},
  title     = {{Optimizing Digital Twin Synchronization in a Finite Horizon}},
  year      = {2022},
  volume    = {},
  number    = {},
  pages     = {2924-2935},
  doi       = {10.1109/WSC57314.2022.10015424}
}

@inproceedings{cakir2023synchdt,
  author    = {Cakir, Lal Verda and Al-Shareeda, Sarah and Oktug, Sema F. and Özdem, Mehmet and Broadbent, Matthew and Canberk, Berk},
  booktitle = {2023 IEEE 28th International Workshop on Computer Aided Modeling and Design of Communication Links and Networks (CAMAD)},
  title     = {{How to synchronize Digital Twins? A Communication Performance Analysis}},
  year      = {2023},
  volume    = {},
  number    = {},
  pages     = {123-127},
  doi       = {10.1109/CAMAD59638.2023.10478422}
}

@article{zheng2023datasync,
  author  = {Zheng, Jinkai and others},
  journal = {IEEE Transactions on Wireless Communications},
  title   = {{Data Synchronization in Vehicular Digital Twin Network: A Game Theoretic Approach}},
  year    = {2023},
  volume  = {22},
  number  = {11},
  pages   = {7635-7647},
  doi     = {10.1109/TWC.2023.3254158}
}

@article{tan2024theDTsynchProblem,
  author    = {Barış Tan and Andrea Matta},
  title     = {{The digital twin synchronization problem: Framework, formulations, and analysis}},
  journal   = {IISE Transactions},
  volume    = {56},
  number    = {6},
  pages     = {652--665},
  year      = {2024},
  publisher = {Taylor \& Francis},
  doi       = {10.1080/24725854.2023.2253869},
  url       = {https://doi.org/10.1080/24725854.2023.2253869},
  eprint    = {https://doi.org/10.1080/24725854.2023.2253869}
}

@inproceedings{munoz2022usingtrace,
  author    = {Mu\~{n}oz, Paula and Wimmer, Manuel and Troya, Javier and Vallecillo, Antonio},
  title     = {{Using trace alignments for measuring the similarity between a physical and its digital twin}},
  year      = {2022},
  isbn      = {9781450394673},
  publisher = {Association for Computing Machinery},
  address   = {New York, NY, USA},
  url       = {https://doi.org/10.1145/3550356.3563135},
  doi       = {10.1145/3550356.3563135},
  booktitle = {Proceedings of the 25th International Conference on Model Driven Engineering Languages and Systems: Companion Proceedings},
  pages     = {503–510},
  numpages  = {8},
  location  = {Montreal, Quebec, Canada},
  series    = {MODELS '22}
}

@article{munoz2024measuring,
  author  = {Muñoz, Paula and Wimmer, Manuel and Troya, Javier and Vallecillo, Antonio},
  journal = {IEEE Transactions on Software Engineering},
  title   = {{Measuring the Fidelity of a Physical and a Digital Twin Using Trace Alignments}},
  year    = {2024},
  volume  = {50},
  number  = {12},
  pages   = {3122-3145},
  doi     = {10.1109/TSE.2024.3462978}
}

@article{salehi2024multiverse,
  author   = {Salehi, Batool and others},
  journal  = {IEEE/ACM Transactions on Networking},
  title    = {{Multiverse at the Edge: Interacting Real World and Digital Twins for Wireless Beamforming}},
  year     = {2024},
  volume   = {32},
  number   = {4},
  pages    = {3092-3110},
  doi      = {10.1109/TNET.2024.3377114}
}

@article{huang2025beam,
  title     = {{Beam management for millimeter-wave mobile communications based on digital twin-enabled scenario cognition}},
  author    = {Huang, Yuhong and Zhao, Youping},
  journal   = {Scientific Reports},
  volume    = {15},
  number    = {1},
  pages     = {13802},
  year      = {2025},
  publisher = {Nature Publishing Group UK London}
}

@article{grieves2014digital,
  title   = {{Digital twin: manufacturing excellence through virtual factory replication}},
  author  = {Grieves, Michael},
  journal = {White paper},
  volume  = {1},
  number  = {2014},
  pages   = {1--7},
  year    = {2014}
}

@article{jones2020characterisingDT,
  title    = {{Characterising the Digital Twin: A systematic literature review}},
  journal  = {CIRP Journal of Manufacturing Science and Technology},
  volume   = {29},
  pages    = {36-52},
  year     = {2020},
  issn     = {1755-5817},
  doi      = {https://doi.org/10.1016/j.cirpj.2020.02.002},
  url      = {https://www.sciencedirect.com/science/article/pii/S1755581720300110},
  author   = {David Jones and Chris Snider and Aydin Nassehi and Jason Yon and Ben Hicks}
}

@techreport{3gpp_tr28915,
  author      = {{3rd Generation Partnership Project (3GPP)}},
  title       = {{Study on management aspect of Network Digital Twin}},
  number      = {TR 28.915},
  institution = {3GPP},
  year        = {2024},
  version     = {18.0.0},
  url         = {https://www.3gpp.org/dynareport/28915.htm}
}

@misc{ETSI-2024-NDT,
  author = {ETSI},
  title  = {{ETSI GR ZSM 015 V1.1.1: Zero-touch network and Service Management (ZSM); Network Digital Twin}},
  year   = {2024}
}

@misc{IETF-2025-arch_proposal,
  author = {IETF},
  title  = {{Network Digital Twin: Concepts and Reference Architecture}},
  year   = {2025}
}

@misc{ORAN-2024-DTRAN_usecases,
  author = {ORAN},
  title  = {{Research Report on Digital Twin RAN Use Cases}},
  year   = {2024}
}

@article{Wu-2021-DT_Networks,
  author  = {Wu, Yiwen and Zhang, Ke and Zhang, Yan},
  journal = {IEEE Internet of Things Journal},
  title   = {{Digital Twin Networks: A Survey}},
  year    = {2021},
  volume  = {8},
  number  = {18},
  pages   = {13789-13804},
  doi     = {10.1109/JIOT.2021.3079510}
}

@article{alghamdi2024synchronization,
  title   = {{Synchronization Patterns for Digital Twin Systems}},
  author  = {Alghamdi, Wael and Albassam, Emad},
  journal = {Journal of Applied Data Sciences},
  volume  = {5},
  number  = {3},
  pages   = {1026--1037},
  year    = {2024}
}

@inproceedings{zhang2018equipment,
  author    = {Zhang, Meng and Zuo, Ying and Tao, Fei},
  booktitle = {2018 IEEE 15th International Conference on Networking, Sensing and Control (ICNSC)},
  title     = {{Equipment energy consumption management in digital twin shop-floor: A framework and potential applications}},
  year      = {2018},
  volume    = {},
  number    = {},
  pages     = {1-5},
  doi       = {10.1109/ICNSC.2018.8361272}
}

@article{luo2019digital,
  title     = {{Digital twin for CNC machine tool: modeling and using strategy}},
  author    = {Luo, Weichao and Hu, Tianliang and Zhang, Chengrui and Wei, Yongli},
  journal   = {Journal of Ambient Intelligence and Humanized Computing},
  volume    = {10},
  number    = {3},
  pages     = {1129--1140},
  year      = {2019},
  publisher = {Springer}
}

@article{zhuang2018digital,
  title     = {{Digital twin-based smart production management and control framework for the complex product assembly shop-floor}},
  author    = {Zhuang, Cunbo and Liu, Jianhua and Xiong, Hui},
  journal   = {The international journal of advanced manufacturing technology},
  volume    = {96},
  number    = {1},
  pages     = {1149--1163},
  year      = {2018},
  publisher = {Springer}
}

@article{zheng2019application,
  title     = {{An application framework of digital twin and its case study}},
  author    = {Zheng, Yu and Yang, Sen and Cheng, Huanchong},
  journal   = {Journal of ambient intelligence and humanized computing},
  volume    = {10},
  number    = {3},
  pages     = {1141--1153},
  year      = {2019},
  publisher = {Springer}
}

@article{guo2019modular,
  title     = {{Modular based flexible digital twin for factory design}},
  author    = {Guo, Jiapeng and Zhao, Ning and Sun, Lin and Zhang, Saipeng},
  journal   = {Journal of Ambient Intelligence and Humanized Computing},
  volume    = {10},
  number    = {3},
  pages     = {1189--1200},
  year      = {2019},
  publisher = {Springer}
}

@inproceedings{charchar2025,
  author    = {Silva et al., Genivaldo},
  booktitle = {{XLIII Brazilian Symposium on Telecommunications and Signal Processing - SBrT 2025}},
  title     = {{Digital Twin of a Beam Selection Procedure in an Indoor Scenario}},
  year      = {2025},
  location  = {Natal, RN, Brazil},
  doi       = {10.14209/sbrt.2025.1571157127},
  url       = {https://doi.org/10.14209/sbrt.2025.1571157127}
}

@article{alzami2025DTindustriesSurvey,
  author   = {Bokhtiar Al Zami, Md and Shaon, Shaba and Khanh Quy, Vu and Nguyen, Dinh C.},
  journal  = {IEEE Access},
  title    = {{Digital Twin in Industries: A Comprehensive Survey}},
  year     = {2025},
  volume   = {13},
  number   = {},
  pages    = {47291-47336},
  doi      = {10.1109/ACCESS.2025.3551532}
}

@techreport{itu2022y3090,
  author      = {{International Telecommunication Union}},
  title       = {{Y.3090: Digital twin network - Requirements and architecture}},
  institution = {{International Telecommunication Union (ITU)}},
  year        = {2022},
  number      = {Y.3090},
  type        = {Recommendation},
  address     = {Geneva, Switzerland},
  url         = {https://www.itu.int/rec/T-REC-Y.3090-202202-I/en},
  urldate     = {2025-07-14},
  note        = {Approved in 2022-02-13}
}

@article{sionna,
  title   = {{Sionna: An Open-Source Library for Next-Generation Physical Layer Research}},
  author  = {Hoydis et al., Jakob},
  year    = {2022},
  month   = {Mar.},
  journal = {arXiv preprint},
  online  = {https://arxiv.org/abs/2203.11854}
}

@book{geron2022hands,
  title     = {{Hands-on machine learning with Scikit-Learn, Keras, and TensorFlow}},
  author    = {G{\'e}ron, Aur{\'e}lien},
  year      = {2022},
  publisher = {" O'Reilly Media, Inc."}
}

@article{2004ColbournCombinatorial,
  author  = {Colbourn, Charles J.},
  title   = {{Combinatorial aspects of covering arrays}},
  journal = {Le Matematiche},
  year    = {2004},
  volume  = {59},
  number  = {1--2},
  pages   = {125--167}
}

@article{2006MooreCombinedArray,
  author  = {Moore, Leslie M. and McKay, Michael D. and Campbell, Katherine S.},
  title   = {{Combined array experiment design}},
  journal = {Reliability Engineering \& System Safety},
  year    = {2006},
  volume  = {91},
  number  = {10},
  pages   = {1281--1289},
  doi     = {10.1016/j.ress.2005.11.024}
}

@article{2012AvilaSimulated,
  author  = {Avila-George, H. and Torres-Jimenez, J. and Hern{\'a}ndez, V. and Gonzalez-Hernandez, L.},
  title   = {{Simulated annealing for constructing mixed covering arrays}},
  journal = {Journal of Applied Research and Technology},
  year    = {2012},
  volume  = {10},
  number  = {4},
  pages   = {658--671}
}

@inproceedings{conceicao2025lowcost,
  author    = {Conceição, Valdinei and others},
  booktitle = {{XLIII Brazilian Symposium on Telecommunications and Signal Processing - SBrT 2025}},
  title     = {{Low-Cost Experimental Platform for AI-Based Beam Management Using WiFi Radio}},
  year      = {2025},
  location  = {Natal, RN, Brazil},
  doi       = {10.14209/sbrt.2025.1571157315},
  url       = {https://doi.org/10.14209/sbrt.2025.1571157315}
}

@misc{valdinei2025repo,
  author       = {Valdinei Conceição},
  title        = {{Samurai mmWaves testbed}},
  howpublished = {\url{https://github.com/Manky0/samurai-servers}},
  year         = {2025},
  note         = {Accessed: 20 August 2025}
}

@misc{WirelessInSiteWebSite,
  title        = {{Wireless EM Propagation Software - Wireless InSite - Remcom}},
  howpublished = {\url{https://www.remcom.com/wireless-insite-em-propagation-software}},
  note         = {Accessed: 2023-11-02}
}

@misc{ITU-2022-DTN_reqs_and_arch,
  author = {ITU},
  title  = {{Y.3090 : Digital twin network - Requirements and architecture}},
  year   = {2022}
}

@inproceedings{salehi2022FLASH,
  author    = {Salehi, Batool and Gu, Jerry and Roy, Debashri and Chowdhury, Kaushik},
  booktitle = {{IEEE INFOCOM 2022 - IEEE Conference on Computer Communications}},
  title     = {{FLASH: Federated Learning for Automated Selection of High-band mmWave Sectors}},
  year      = {2022},
  volume    = {},
  number    = {},
  pages     = {1719-1728},
  doi       = {10.1109/INFOCOM48880.2022.9796865}
}

@inproceedings{bergstra2011algorithms,
  author    = {Bergstra, James and Bardenet, R\'{e}mi and Bengio, Yoshua and K\'{e}gl, Bal\'{a}zs},
  booktitle = {Advances in Neural Information Processing Systems},
  editor    = {J. Shawe-Taylor and R. Zemel and P. Bartlett and F. Pereira and K.Q. Weinberger},
  pages     = {},
  publisher = {Curran Associates, Inc.},
  title     = {{Algorithms for Hyper-Parameter Optimization}},
  url       = {https://proceedings.neurips.cc/paper_files/paper/2011/file/86e8f7ab32cfd12577bc2619bc635690-Paper.pdf},
  volume    = {24},
  year      = {2011}
}

\end{document}